\documentclass[prl,aps,floatfix,twocolumn,showpacs,nofootinbib,superscriptaddress]{revtex4}

\usepackage{amsmath}
\usepackage{amssymb}
\usepackage{amsfonts}
\usepackage{graphicx}
\usepackage{color}

\allowdisplaybreaks

\newcommand{\nuc}[2]{$^{#1}${#2}}

\begin{document}

%\title{Generator Coordinate Method based on
%       angular-momentum and particle-number projected
%       blocked triaxial one-quasiparticle HFB states for the
%       odd-$A$ nucleus \nuc{25}{Mg}}

\title{Beyond Mean-Field Calculations for Odd-Mass Nuclei}

\author{B. Bally}
\affiliation{Universit{\'e} de Bordeaux,
             Centre d'Etudes Nucl{\'e}aires de Bordeaux Gradignan,
             Chemin du Solarium, BP120, F-33175 Gradignan, France}
\affiliation{CNRS/IN2P3,
             Centre d'Etudes Nucl{\'e}aires de Bordeaux Gradignan,
             Chemin du Solarium, BP120, F-33175 Gradignan, France}

\author{B. Avez}
\affiliation{Universit{\'e} de Bordeaux,
             Centre d'Etudes Nucl{\'e}aires de Bordeaux Gradignan,
             Chemin du Solarium, BP120, F-33175 Gradignan, France}
\affiliation{CNRS/IN2P3,
             Centre d'Etudes Nucl{\'e}aires de Bordeaux Gradignan,
             Chemin du Solarium, BP120, F-33175 Gradignan, France}

\author{M. Bender}
\affiliation{Universit{\'e} de Bordeaux,
             Centre d'Etudes Nucl{\'e}aires de Bordeaux Gradignan,
             Chemin du Solarium, BP120, F-33175 Gradignan, France}
\affiliation{CNRS/IN2P3,
             Centre d'Etudes Nucl{\'e}aires de Bordeaux Gradignan,
             Chemin du Solarium, BP120, F-33175 Gradignan, France}

\author{P.-H. Heenen}
\affiliation{PNTPM, CP229,
             Universit{\'e} Libre de Bruxelles,
             B-1050 Bruxelles,
             Belgium}
%
%=======================================================================
%

\begin{abstract}

Beyond mean-field methods are very successful tools for the description of large-amplitude collective motion for even-even atomic nuclei. 
The state-of-the-art framework of these methods consists in a Generator Coordinate Method based on angular-momentum and particle-number projected triaxially deformed Hatree-Fock-Bogoliubov (HFB) states. 
The extension of this scheme to odd-mass nuclei is a long-standing challenge.
We present for the first time such an extension, where
the Generator Coordinate space is built from self-consistently blocked one-quasiparticle HFB states.
One of the key points for this success is that the same Skyrme interaction is used for the mean-field and the pairing channels, thus avoiding problems related to the violation of the Pauli principle.
An application to \nuc{25}{Mg}
illustrates the power of our method, as agreement with experiment is obtained for the spectrum, electromagnetic moments, and transition strengths, for both positive and negative parity states and without the necessity for effective charges or effective moments. Although the effective interaction still requires improvement, our study opens the way to systematically describe odd-$A$ nuclei throughout the nuclear chart.

\end{abstract}


\date{8 October 2014}

\maketitle
%
%=======================================================================
%

Atomic nuclei are a prime example of finite-size, self-bound quantal
many-body systems. Their complex spectra exhibit a large
variety of excitation modes that can be related to either collective or
single-particle degrees of freedom, or to coupling between
both~\cite{Rowe70a,Rin80a}. The symmetries of the Hamiltonian
and the related quantum numbers, chiefly angular momentum and parity,
are the means by which to classify and interpret the energy levels
and the transition probabilities between them.

To go beyond simple models requires the modeling of the
in-medium nucleon-nucleon interaction. We focus here on methods based
on an energy density functional (EDF), which are widely used for
the description of atomic nuclei \cite{RMP} and electronic systems
\cite{Cluster}.
Their simplest realization is the self-consistent mean-field (SCMF) method
along the lines of the Hartree-Fock (HF) and Hartree-Fock-Bogoliubov (HFB)
schemes. In a SCMF approach, correlations related to shape deformation and
pairing are incorporated at a moderate numerical cost, but at the
price of breaking symmetries of the nuclear Hamiltonian. Such breakings
prevent a detailed comparison with experimental data. In particular,
transition probabilities between levels can only be estimated when making
the additional assumptions of the collective model~\cite{Rowe70a,Rin80a}.

To go a step further requires a so-called "beyond mean-field model",
while taking into account correlations
absent in the SCMF. Two such extensions are the restoration of symmetries
broken by the mean field and the superposition of different configurations
by the Generator Coordinate Method (GCM). Such a multireference (MR)
approach is particularly well suited to describe shape coexistence and
shape mixing phenomena. As evidenced by many applications to even-even
nuclei, this
family of methods describes well a large range of nuclear properties
\cite{Val00a,Rod02b,Nik06a,Ben08a,Yao10a,Rod10a,Ben06a}.
These extensions were until now limited to the study of even-even
nuclei. Already at the SCMF level, the description of odd-$A$ nuclei
\cite{duguet02a,Ber09a,Schu10a,pot13a,tar14a} poses new difficulties.
Breaking a nucleon pair unavoidably lifts the time-reversal symmetry
of the HFB state, and several low-lying blocked HFB
states usually lie close in energy and have to be calculated separately
in a fully self-consistent manner to determine the level ordering.

Here, we present the first results obtained with a gene\-ra\-lization of our
method for even-even nuclei \cite{Ben08a} to odd ones. The MR basis
is constructed from angular-momentum projected (AMP) and
particle-number projected (PNP) self-consistently blocked triaxial
one-quasiparticle (1qp) HFB states. Having to consider several blocked states
at each deformation makes the calculation much larger than state-of-the-art
ones for even-even nuclei. The breaking of time-reversal symmetry has
important practical consequences. One is that it makes the formal problems
\cite{anguiano01a,dobaczewski07a,lacroix09a,bender09a} associated with
defining the nondiagonal energy kernels for standard parametrizations
of the EDF even more acute than in the case of even-even nuclei.
These problems are related to a violation of the Pauli principle when
constructing the EDF, and can be avoided by using the same
(density-independent) Skyrme Hamiltonian as particle-hole and pairing
forces. For this purpose, we use the recent SLyMR0 parametrization
\cite{sadoudi13a}.

We have chosen \nuc{25}{Mg} as an example, i.e.\ a light deformed nucleus
exhibiting coexisting rotational bands of both parities
at low excitation energies \cite{Firestone09a}, with band heads
interpreted as one-quasiparticle states \cite{bandheads}.
The adjacent even \nuc{24}{Mg} nucleus has been the testing ground
for many implementations of the MR EDF method for even-even
nuclei \cite{Val00a,Rod02b,Nik06a,Ben08a,Yao10a,Rod10a}.

In the past, angular-momentum projection for odd-$A$ nuclei has been mostly
performed on HF or HFB states constructed in small valence spaces
\cite{Bas65a,Gun67a,Rath93a,Har84a,Ham85a}.
A GCM mixing based on parity and angular-momentum projected
symmetry-unrestricted Slater determinants in a model space of antisymmetrized
Gaussian wave packets has been carried out in the frameworks of
Antisymmetrized~\cite{kanada10a,kimura13a} and Fermionic~\cite{neff08a}
Molecular Dynamics.

Our method can be divided into four successive steps. First, a set of
"false HFB vacua" \cite{bender00a,duguet02a} is generated, consisting of fully
paired and time-reversal-invariance-conserving nonblocked HFB states
constrained to particle numbers \mbox{$Z=12$} and \mbox{$N=13$}. In its
canonical basis, each is given by
$| \text{HFB}_{\text{fv}} (q_1,q_2) \rangle
= \prod_{k > 0} ( u_k + v_k a^\dagger_{k} a^\dagger_{\bar{k}} ) | - \rangle
$, where the single-particle states are chosen to conserve three point-group
symmetries \cite{dob00a}, namely parity $\pi$, a signature, and a time
simplex, which leads to nucleon densities with triaxial
symmetry \cite{Hel12a}. Thanks to a
constraint on the mass quadrupole moment added to the HFB equations and
parame\-terized by $q_1$ and $q_2$ as defined in \cite{Ben08a}, one
sextant of the $\beta$-$\gamma$ plane with $0 \leq \gamma \leq 60^\circ$
is covered.

These vacua, however, only serve to identify the single-particle states
next to the Fermi energy at each deformation. In a second step,
several one-quasiparticle HFB states are then constructed from
each false vacuum by self-consistently blocking the most favored
configurations with the method described in Ref.~\cite{gall94a}.
In its respective canonical basis, each has the structure
$| \text{HFB}^\pi_{\text{1qp}} (q_1,q_2,j) \rangle
= a^\dagger_{j} \prod_{k \neq j > 0 }
( u_k + v_k a^\dagger_{k} a^\dagger_{\bar{k}} ) | - \rangle
$, and therefore each adopts the parity and signature of the blocked
single-particle state $j$.
Whenever possible we use the compact label $\mu \equiv q_1, q_2, j$ to
distinguish between 1qp states $| \text{HFB}^\pi_{\text{1qp}} (\mu) \rangle$
that differ in any of the three coordinates.
There are two other nonequivalent sextants of the $\beta$-$\gamma$
plane that will not be considered here. For those, the conserved signature
is aligned with a different major axis of the quadrupole tensor of the nucleus,
leading to a slightly different total energy of the 1qp states \cite{Schu10a}.
We also omit 3qp and higher multiquasiparticle states.

The 1qp states break several symmetries of the nuclear Hamiltonian.
The third step of our method restores the most important ones for
nuclear spectroscopy applications: the proton $(Z)$ and neutron $(N)$ numbers,
and the angular momentum $\hbar^2 \, J (J+1)$ with $z$ component $\hbar M$,
\begin{equation}
\label{eq:proj}
| J^\pi M \kappa (\mu) \rangle
= \sum_{K=-J}^{J} f^{J^\pi \kappa}_{\mu,K} \,
  \hat{P}^{J}_{M K} \hat{P}^{N} \hat{P}^{Z}
  | \text{HFB}^\pi_{\text{1qp}}(\mu) \rangle
\, .
\end{equation}
The indices for $N$ and $Z$ are dropped from
$| J^\pi M \kappa \rangle$ as all states are projected on
$N=13$ and $Z=12$.
The 1qp states are developed into angular momentum eigenstates with $z$
component $\hbar K$. Their weights $f^{J^\pi \kappa}_{\mu,K}$ are
determined by solving a Hill-Wheeler-Griffin (HWG) equation
\cite{BlaRip,Rin80a}
\begin{equation}
\label{eq:hwK}
\sum_{K'}
\Big( \mathcal{H}^{J^\pi}_{\mu,K; \mu, K'}
      - E^{J^\pi}_{\kappa} \mathcal{I}^{J^\pi}_{\mu,K; \mu, K'}
\Big) \, f^{J^\pi \kappa}_{\mu,K'}
 = 0
\end{equation}
for each $J$ on which the 1qp state can
be projected, where $\mathcal{H}^{J^\pi}_{\mu, K; \mu', K'}
\equiv \langle \text{HFB}^\pi_{\text{1qp}}(\mu) | \hat{H} \hat{P}^{J}_{K K'}
\hat{P}^Z \hat{P}^N | \text{HFB}^\pi_{\text{1qp}}(\mu') \rangle$ and
$\mathcal{I}^{J^\pi}_{\mu,K; \mu', K'}
\equiv \langle \text{HFB}^\pi_{\text{1qp}}(\mu) | \hat{P}^J_{K K'} \hat{P}^Z
\hat{P}^N | \text{HFB}^\pi_{\text{1qp}}(\mu') \rangle$ are the Hamiltonian
and norm kernels, respectively. As a consequence of the signature
symmetry of the $| \text{HFB}^\pi_{\text{1qp}} (\mu) \rangle$, components
with $\pm K$ are linearly dependent. The redundant ones are removed
by a transformation, as proposed in Ref.~\cite{Ena99a}.
For each value of $J$, one obtains in this way a spectrum of up to $(2J+1)/2$
states of energy $E^{J^\pi}_{\kappa}$ labeled by an index $\kappa$.

In the final step, projected states obtained from different 1qp states
are mixed by the GCM
\begin{equation}
\label{eq:gcmanss}
| J^\pi M \xi \rangle
= \sum_{\mu=1}^{\Omega_\pi}
  \sum_{\kappa} f^{J^\pi \xi}_{\mu,\kappa} \, | J^\pi M \kappa (\mu) \rangle
\, ,
\end{equation}
where $\Omega_\pi$ is the number of different 1qp states $ | \text{HFB}^\pi_{\text{1qp}}(\mu) \rangle$ 
of given parity $\pi$ that are projected.

The weights $f^{J^\pi \xi}_{\mu,\kappa}$ are determined by a HWG equation
similar to Eq.~\eqref{eq:hwK}, where the energy and norm kernels are now
calculated using the $| J^\pi M \kappa' (\mu') \rangle$ states, and where
the $E^{J^\pi}_{\xi}$ are the energies of the mixed projected
states \eqref{eq:gcmanss}.
As the Hamiltonian commutes with parity, the energy kernels between
states of opposite parity are zero, such that they do not mix in
the GCM. For each value of $J$, the HWG equation is thus solved
separately for positive- and negative-parity states. Having determined
the $f^{J^\pi \kappa}_{\mu,K}$ and $f^{J^\pi \xi}_{\mu,\kappa}$ coefficients,
other observables can be computed as well~\cite{Rod02b,Nik06a,Ben08a}.

The single-particle states are discretized on a Cartesian coordinate-space
mesh in a 3D box. The mean-field calculations are performed
using an update of the code described in \cite{gall94a,Hel12a}.
The projection operators involve rotations and
integrations over gauge angles for PNP and Euler angles for AMP.
These are discretized with 9 points in the interval
$[0, \pi]$ for PNP for protons and neutrons separately, and
$24 \times 40 \times 24$ points for the Euler angles in the full integration
interval $\alpha \in [0,2\pi]$, $\beta \in [0,\pi]$, $\gamma \in [0,2\pi]$.
The remaining symmetries of the 1qp states allow for a reduction to 1/16
of the number of spatial rotations to be explicitly carried out.
When calculating the GCM kernels, derivatives and the spatial rotations
are carried out with Lagrange-mesh techniques \cite{baye86a}.

\begin{figure}[t!]
\begin{minipage}[t]{1.0cm} \vspace*{-5.21cm} %\vspace*{-2.0cm}
  \centering \includegraphics[width=0.8cm]{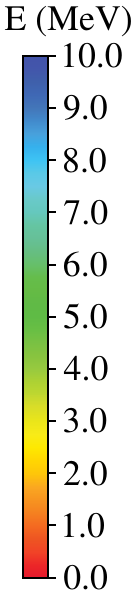}
\end{minipage}%
\begin{minipage}[c]{6.8cm}
\centerline{
  \includegraphics[width=3.4cm]{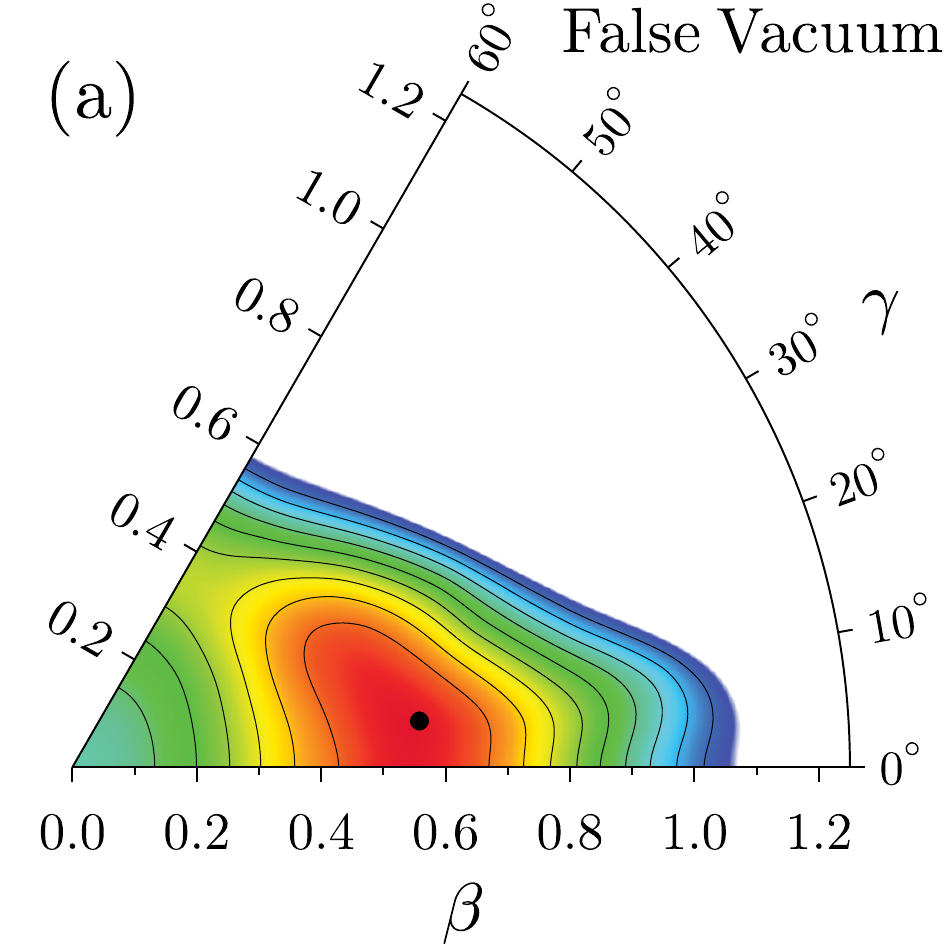}
  \includegraphics[width=3.4cm]{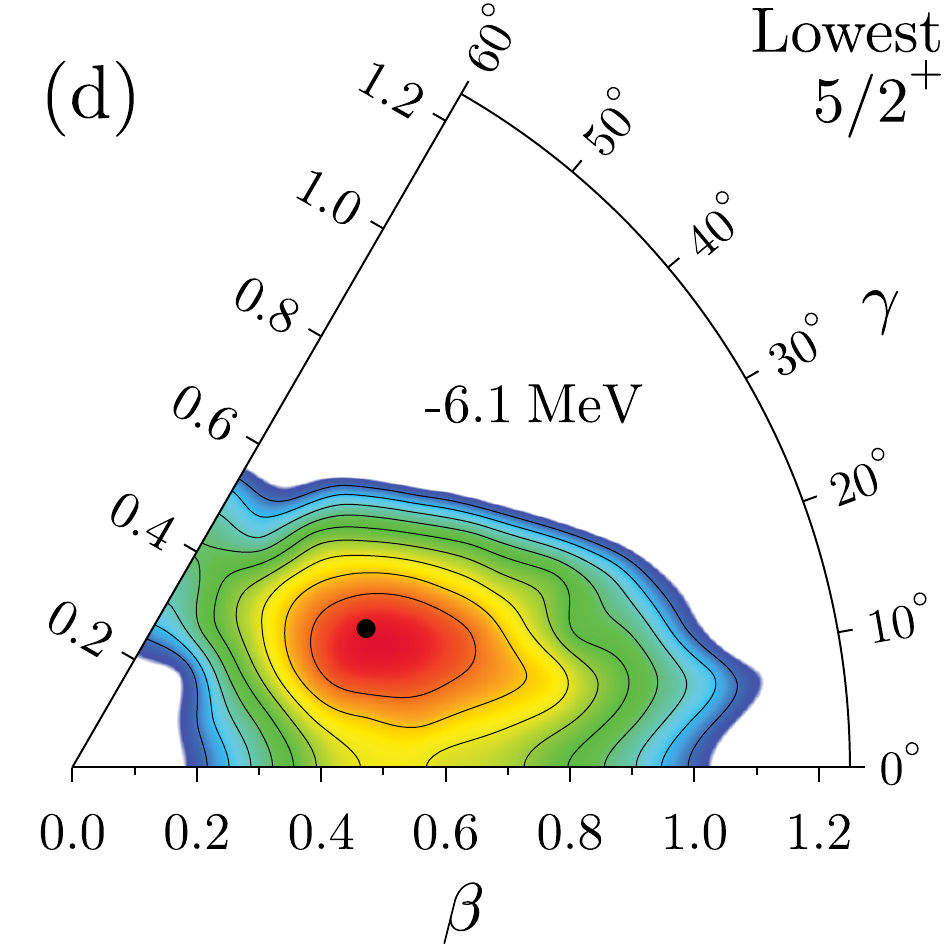}
}
\centerline{
  \includegraphics[width=3.4cm]{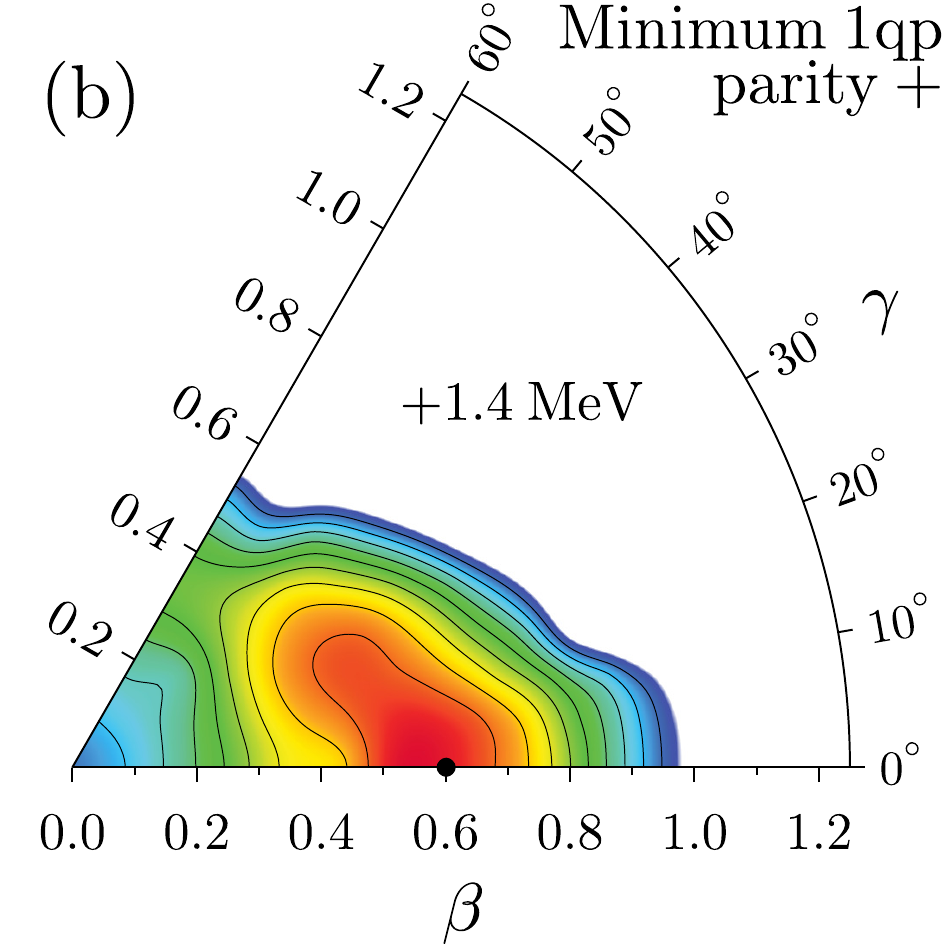}
  \includegraphics[width=3.4cm]{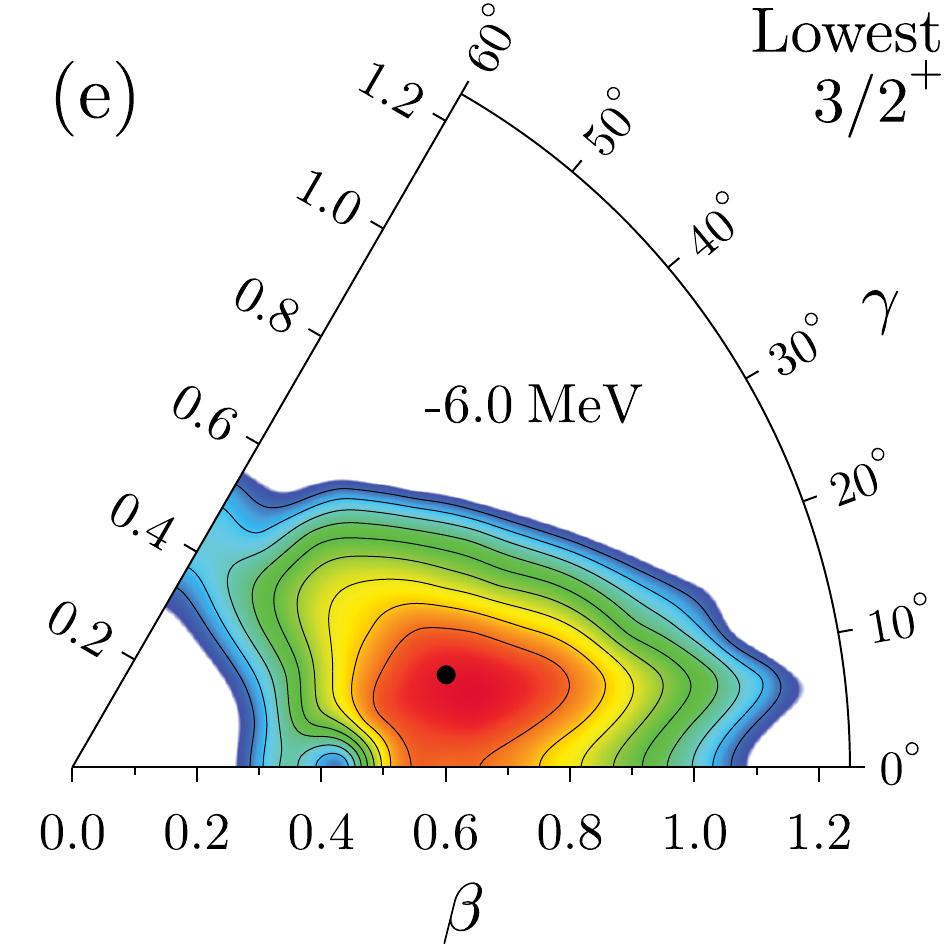}
}
\centerline{
  \includegraphics[width=3.4cm]{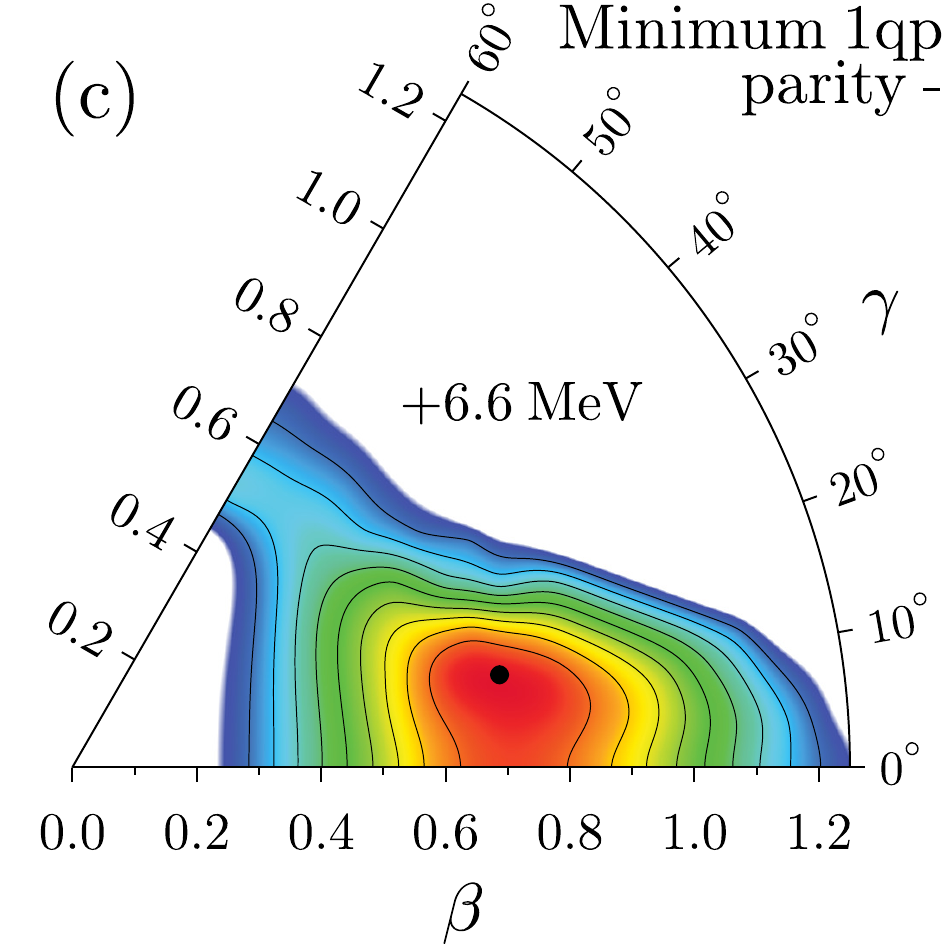}
  \includegraphics[width=3.4cm]{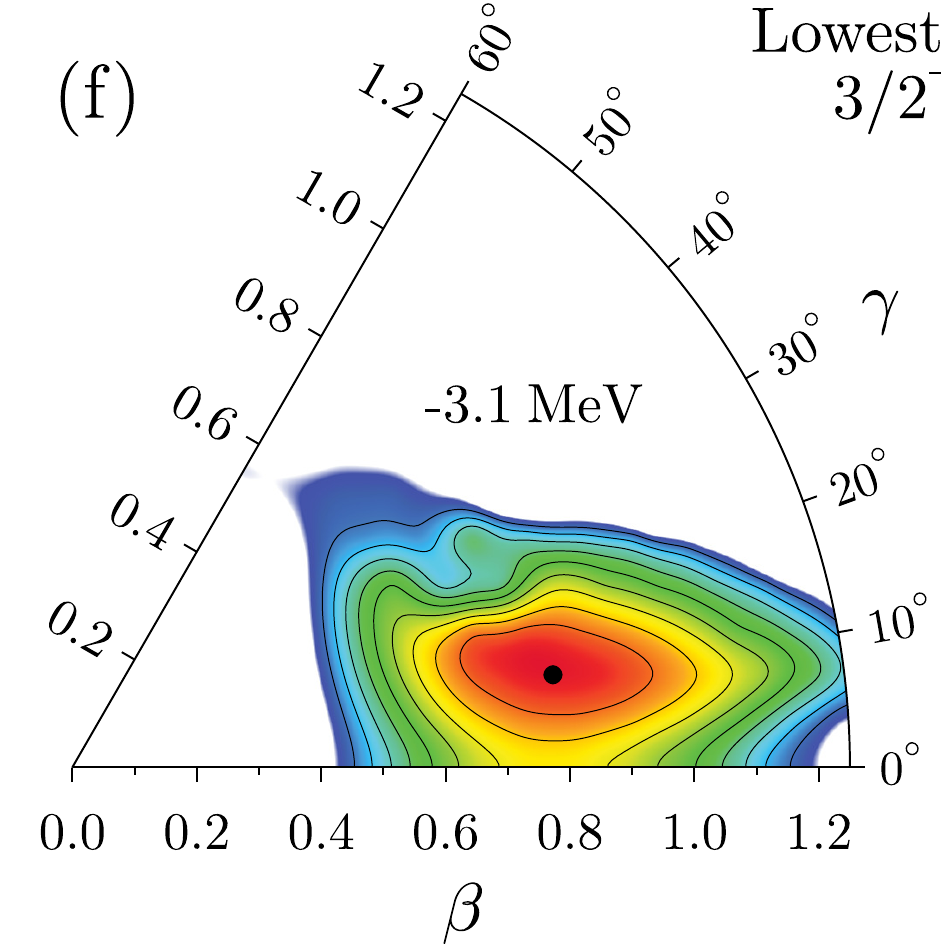}
}
\end{minipage}%

\caption{
\label{fig:surf}
(color online) Energy surface of the false vacuum (a),
surfaces of the lowest energy found at a given deformation among the
several nonprojected 1qp states of positive (b), and negative (c) parity,
and surfaces of the lowest energy found at a given deformation after
projection on $N$, $Z$, and $J^\pi = 5/2^+$ (d), $3/2^+$ (e), and
$3/2^-$ (f) and $K$ mixing among the 1qp states.
All energies are displayed as a function of $\beta$ and $\gamma$
calculated from the mass density distribution of the 1qp state as
defined in Ref.~\cite{Ben08a}.
The deformation energy is relative to the minimum of each
surface, indicated by a black dot.
The offset of the minimum of each surface relative to the one of
false vacua is indicated in the upper corner of each sextant.
}
\end{figure}

As mentioned above, the MR calculation described here could not be safely
carried out with a standard Skyrme EDF because of the nondiagonal energy
kernels $\mathcal{H}^{J^\pi}_{\mu,K; \mu', K'}$ becoming ill defined
\cite{anguiano01a,dobaczewski07a,lacroix09a,bender09a}.
Instead, the energy has to be calculated as the matrix element of a
(nondensity-dependent) many-body Hamiltonian without any
approximation or simplification.
The early parametrization SV, used along these lines
in Refs.~\cite{satula10a,satula12a}, is repulsive in the
pairing channel~\cite{sadoudi13a}. Here, we use the Skyrme
parametrization SLyMR0 \cite{sadoudi13a}.
It acts both as a particle-hole and a pairing force.
SLyMR0 consists of standard central and spin-orbit two-body terms with
gradients that are supplemented by gradientless three-body and four-body
terms. Its parameters have been adjusted to provide attractive pairing
of a reasonable size and to avoid instabilities in all spin-isospin channels
\cite{sadoudi13a}. Being overconstrained by these two conditions,
its overall predictive power for nuclear bulk properties is
limited~\cite{sadoudi13a}. Also, its very low isoscalar effective
mass of \mbox{$m^\ast_0/m = 0.47$} leads to a single-particle spectrum
that is too spread out.
The Coulomb energy kernels for the GCM are also calculated with exact
exchange and pairing contributions. In the HFB calculations, however, where
preserving the Pauli principle is less critical, the Slater approximation
is used for the exchange term and the Coulomb pairing energy is neglected.
We use a soft pairing cutoff when solving the HFB equations \cite{gall94a}
in order to suppress the divergence of the HFB equations when using contact interactions \cite{Yu03a}.
By contrast, we omit such cutoff when calculating the GCM energy kernels, as it would introduce
a slight violation of the Pauli principle and thereby lead to the problems with nondiagonal energy kernels 
discussed in \cite{anguiano01a,dobaczewski07a,lacroix09a,bender09a}.
The HFB calculations are augmented by a Lipkin-Nogami scheme that enforces
the presence of pair correlations in most 1qp states.

The Hamiltonian and other operator kernels are evaluated with the
technique presented in Refs.~\cite{Bon90a,Hee93a}. The
corresponding overlap kernels, including their sign, are calculated
with the Pfaffian-based expression of Ref.~\cite{avez12a}.

%
%===================================================================
%

The energy surfaces corresponding to the first steps of our calculations
are plotted in Fig.~\ref{fig:surf}. Figure 1(a) corresponds to the false
vacuum of \nuc{25}{Mg}. The surface is very similar to the one obtained
from HFB calculations for \nuc{24}{Mg}, with a minimum corresponding to a well-deformed,
slightly triaxial, prolate shape. Figure 1(b) and (c) display
the surfaces corresponding to the lowest nonprojected 1qp configurations
for positive and negative parity, respectively. The configuration giving
the lowest energy is selected for each deformation. Compared to panel~(a),
the minima are shifted, which reflects how the blocked single-particle levels
approach and depart from the Fermi energy for neutrons. Figure 1(d) corresponds
to the third step of our method for \mbox{$J^\pi = 5/2^+$}. The surface
is formed by the $K$-mixed states $| J^\pi M \kappa (q_1,q_2,j) \rangle$
projected on $N$ and $Z$, with the lowest energy for a given intrinsic
deformation $(q_1,q_2)$, respectively. Figure 1(e) and (f) display the same result for
\mbox{$J^\pi = 3/2^+$} and $3/2^-$.
The deformation corresponding to the lowest energy is different for
most $J^\pi$ values, and it does not coincide with the one of
the lowest nonprojected blocked 1qp state.
As found in similar calculations for light even-even nuclei
\cite{Ben08a,Yao10a,Rod10a}, AMP shifts the minimum to larger
intrinsic deformation.

\begin{figure}[t!]
 \centerline{\includegraphics[width=8.5cm]{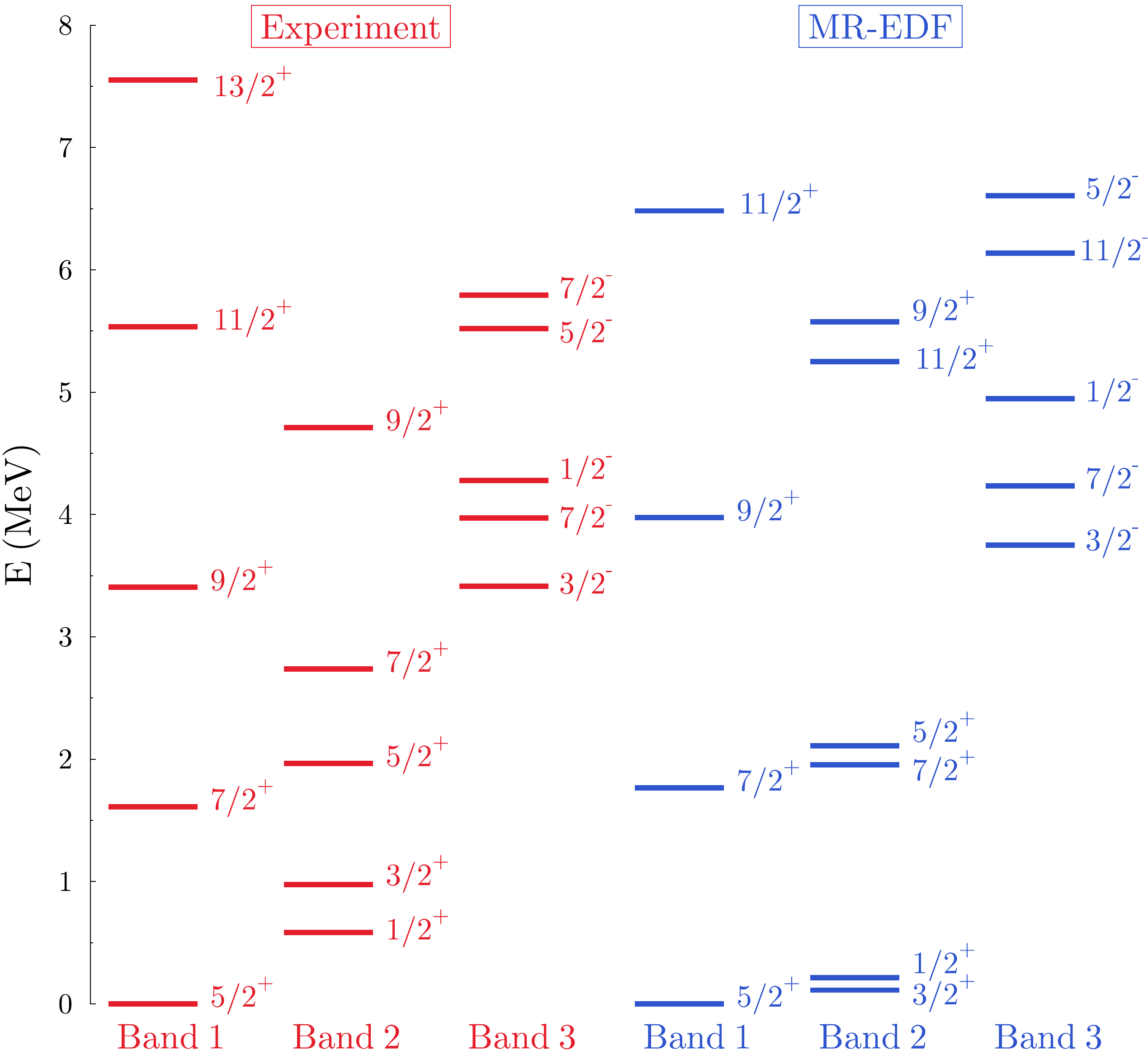}}
\caption{
\label{fig:bands}
(color online)
Comparison of the calculated low-lying levels grouped into rotational bands
with data taken from Ref.~\cite{Firestone09a}.
}
\end{figure}

Let us now discuss the full calculation where the projected states $| J^\pi M \kappa (q_1,q_2,j) \rangle$
with the same value of $J^\pi$ and $M$, but different values for $(q_1,q_2)$, $j$, or $\kappa$, are combined for a
 complete GCM calculation.
 Sampling the deformations $q_1$ and $q_2$ in
steps of $40 \, \text{fm}^2$ and considering several 1qp states
at each combination, we have constructed a basis of
$\Omega_+ = 100$ 1qp configurations of positive parity and
$\Omega_- = 60$ 1qp configurations of negative parity, respectively.
After elimination of all redundant states, the Hamiltonian is
finally diagonalized in a space of 226 $K$-mixed projected states for
$J^\pi = 5/2^+$, 149 for $J^\pi = 3/2^+$, and 106 for $J^\pi = 3/2^-$,
to give a few examples. The large number of 1qp configurations makes
it possible to analyze the calculation's convergence. When adding
the states to the GCM in the order of the energy of the 1qp state they
are projected from, the last 20 states being considered for each parity
only add about 20~keV to the energies of the low-lying states, with their
energy differences changing even less. This smooth convergence could
only be achieved by using a Hamiltonian in the GCM.

\begin{figure}[t!]
\centerline{\includegraphics[height=7.0cm]{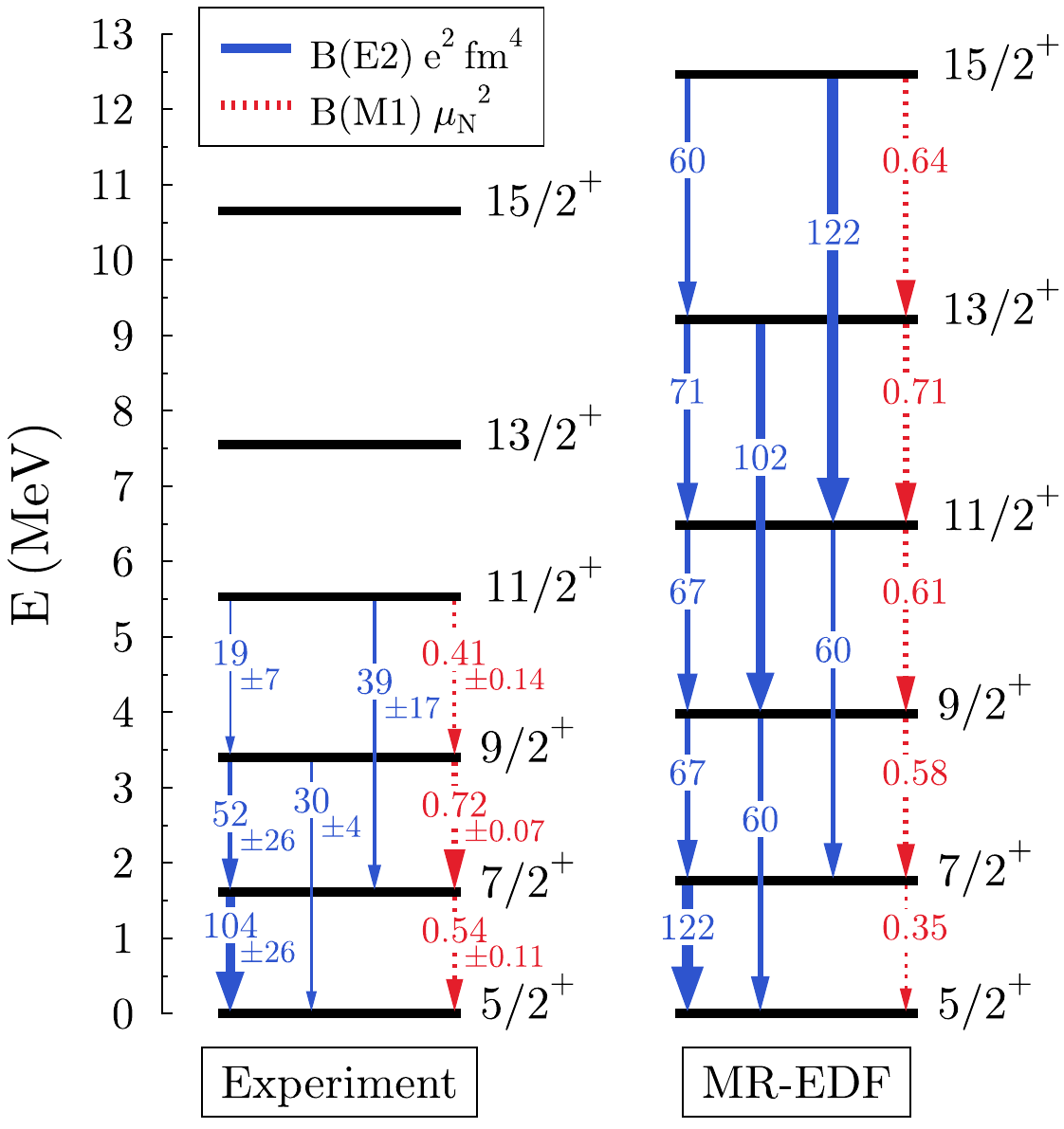}}
\caption{
\label{fig:gsband}
(color online)
Excitation energies of states in the ground-state band of \nuc{25}{Mg}
and $B(E2)$ and $B(M1)$ values for transitions between
them.
}
\end{figure}

The GCM gives a quite satisfying description of the low-lying
levels, see Fig.~\ref{fig:bands}: the overall band structure is
reasonably well reproduced, including the excitation energy of the lowest
levels with negative parity. Band 2, however, has an incorrect signature
splitting and its band head is computed somewhat too low in energy.
Within each band, the spectrum is slightly too spread out, as is also
found for even-even nuclei. This can be corrected for by projecting HFB
states cranked to finite angular momenta; this will be discussed
elsewhere. Computed moments $23.25 \, e \, \text{fm}^2$ and
$-1.054 \, \mu_N$ reproduce the experimental
values of $20.1(3) \, e \, \text{fm}^2$ \cite{Firestone09a} for the
spectroscopic quadrupole moment and $-0.85545(8) \, \mu_N$
\cite{Firestone09a} for the magnetic moment of the ground state
reasonably well using the bare charges and magnetic moments for
the nucleons. The $B(E2)$ and $B(M1)$ values for
transitions within the ground-state band are similarly well described,
see Fig.~\ref{fig:gsband}, with the $B(E2)$ values being
again slightly overestimated. We attribute this to a
single-particle spectrum for SLyMR0 that is too spread out, pushing the dominant intrinsic
configurations to slightly too-large deformations.

Despite its deficiencies for bulk properties such as
masses \cite{sadoudi13a}, SLyMR0 gives a very reasonable description
of the spectroscopy of \nuc{25}{Mg}. Still, there is an urgent
need for effective Hamiltonians that attain at least the predictive
power of the current standard EDFs. Their construction will require
higher-order terms in the effective interaction \cite{sadoudi13b,raimondi14a}.

The beyond mean-field method described here will be a useful tool
to study ground-state correlations and spectroscopy of odd-$A$ nuclei
But let us also stress that the method that we have introduced has an interest beyond the study
of odd nuclei. 
The ability to study isotopic (isotonic) chains including all numbers of neutrons (protons) 
will enlarge the perspective on the systematics of global nuclear properties. Examples of particular current interest 
concern the evolution with nucleon number of signatures of shell effects and pairing correlations.
Examples concern the evolution of signatures of shell effects and pairing correlations with nucleon
number.

%
%===================================================================
%

\textit{Acknowledgements.}
We thank R.~Janssens for a critical reading of the manuscript.
This research has been supported in parts by the
Agence Nationale de la Recherche under Grant No.\ ANR 2010
BLANC 0407 "NESQ",
by the CNRS/IN2P3 through PICS No.\ 5994, and by the PAI-P6-23 of
the Belgian Office for Scientific Policy.
The computations were performed using HPC resources from GENCI-IDRIS
(Grants No.\ 2012-050707 and 2013-050707) and of the MCIA (M{\'e}socentre
de Calcul Intensif Aquitain) of the Universit{\'e} de Bordeaux and of
the Universit{\'e} de Pau et des Pays de l'Adour.

%
%===================================================================
%


\begin{thebibliography}{99}

\bibitem{Rowe70a}
D. J. Rowe,
\textit{Nuclear Collective Motion}
(Methuen, London, 1970).

\bibitem{Rin80a}
P. Ring and P. Schuck,
\textit{The Nuclear Many-Body Problem}
(Springer, New York, Heidelberg, Berlin, 1980).

\bibitem{RMP}
M. Bender, P.-H. Heenen, and P.-G. Reinhard,
Rev. Mod. Phys. \textbf{75}, 121 (2003).

\bibitem{Cluster}
P.-G. Reinhard and E. Suraud,
\textit{Introduction to Cluster Dynamics}
(Wiley-VCH, Berlin, 2003).

\bibitem{Val00a}
A. Valor, P.-H. Heenen, and P. Bonche,
Nucl. Phys. \textbf{A671}, 145 (2000).

\bibitem{Rod02b}
R. R. Rodr{\'i}guez-Guzm{\'a}n, J. L. Egido, and L. M. Robledo,
Nucl. Phys. \textbf{A709}, 201 (2002).

\bibitem{Nik06a}
T. Nik{\v{s}}i{\'c}, D. Vretenar, and P. Ring,
Phys. Rev. C \textbf{74}, 064309 (2006).

\bibitem{Ben08a}
M. Bender and P.-H. Heenen,
Phys. Rev. C \textbf{78}, 024309 (2008).

\bibitem{Yao10a}
J. M. Yao, J. Meng, P. Ring, and D. Vretenar,
Phys. Rev. C \textbf{81}, 044311 (2010).

\bibitem{Rod10a}
T. R. Rodr{\'i}guez and J. L. Egido,
Phys. Rev. C \textbf{81}, 064323 (2010).

\bibitem{Ben06a}
M. Bender, G. F. Bertsch and P.-H. Heenen,
Phys. Rev. C \textbf{73}, 034322 (2006).

\bibitem{duguet02a}
T. Duguet, P. Bonche, P.-H. Heenen, and J. Meyer,
Phys. Rev. C \textbf{65}, 014310 (2001).

\bibitem{Ber09a}
G. Bertsch, J. Dobaczewski, W. Nazarewicz, and J. Pei,
Phys. Rev. A \textbf{79}, 043602 (2009).

\bibitem{Schu10a}
N. Schunck, J. Dobaczewski, J. McDonnell, J. Mor{\'e},
W. Nazarewicz, J. Sarich, and M. V. Stoitsov,
Phys. Rev. C \textbf{81}, 024316 (2010).

\bibitem{pot13a}
K. J. Pototzky, J. Erler, P.-G. Reinhard, and V. O. Nesterenko,
Eur. Phys. J. A \textbf{46}, 299 (2010).

\bibitem{tar14a}
D. Tarpanov, J. Toivanen, J. Dobaczewski, and B. G. Carlsson,
Phys. Rev. C \textbf{89}, 014307 (2014).

\bibitem{anguiano01a}
M. Anguiano, J. L. Egido, and L. M. Robledo,
Nucl. Phys. \textbf{A696}, 467 (2001).

\bibitem{dobaczewski07a}
J. Dobaczewski, M. V. Stoitsov, W. Nazarewicz, and P.-G. Reinhard,
Phys. Rev. C \textbf{76}, 054315 (2007).

\bibitem{lacroix09a}
D. Lacroix, T. Duguet, and M. Bender,
Phys. Rev. C \textbf{79}, 044318 (2009).

\bibitem{bender09a}
M. Bender, T. Duguet, and D. Lacroix,
Phys. Rev. C \textbf{79}, 044319 (2009).

\bibitem{sadoudi13a}
J. Sadoudi, M. Bender, K. Bennaceur, D. Davesne, R. Jodon, and T. Duguet,
Phys. Scr. \textbf{T154}, 014013 (2013).

\bibitem{Firestone09a}
R. B. Firestone,
Nuclear Data Sheets \textbf{110}, 1691 (2009).

\bibitem{bandheads}
Y. Fujita, I. Hamamoto, H. Fujita, Y. Shimbara, T. Adachi, G. P. A. Berg,
K. Fujita, K. Hatanaka, J. Kamiya, K. Nakanishi, Y. Sakemi, Y. Shimizu,
M. Uchida, T. Wakasa, and M. Yosoi,
Phys. Rev. Lett. \textbf{92}, 062502 (2004).

\bibitem{Bas65a}
W. H. Bassichis, B. Giraud, and G. Ripka,
Phys. Rev. Lett. \textbf{15}, 980 (1965).

\bibitem{Gun67a}
M. R. Gunye and C. S. Warke,
Phys. Rev. \textbf{156}, 1087 (1967).

\bibitem{Rath93a}
A. K. Rath, C. R. Praharaj, and S. B. Khadkikar,
Phys. Rev. C \textbf{47}, 1990 (1993).

\bibitem{Har84a}
K. Hara and S. Iwasaki,
Nucl. Phys. \textbf{A430}, 175 (1984).

\bibitem{Ham85a}
E. Hammar{\'e}n, K. W. Schmid, F. Gr{\"u}mmer, A. Faessler, and B. Fladt,
Nucl. Phys. \textbf{A437}, 1 (1985).

\bibitem{kimura13a}
M. Kimura, Y. Taniguchi, Y. Kanada-En'yo, H. Horiuchi, and K. Ikeda,
Phys. Rev. C \textbf{87}, 011301 (2013).

\bibitem{kanada10a}
Y. Kanada-En'yo and M. Kimura,
in
C. Beck (ed.), \textit{Clusters in Nuclei},
Lecture Notes in Physics Vol. \textbf{818}
(Springer Verlag, Heidelberg, 2010).

\bibitem{neff08a}
T. Neff and H. Feldmeier,
Eur. Phys. J. Spectial Topics \textbf{156}, 69 (2008).

\bibitem{bender00a}
M. Bender, K. Rutz, P.-G. Reinhard, and J. A. Maruhn,
Eur. Phys. J. A \textbf{8}, 59 (2000).

\bibitem{dob00a}
J. Dobaczewski, J. Dudek, S. G. Rohozi{\'n}ski, and T. R. Werner,
Phys. Rev. C \textbf{62}, 014310 (2000).

\bibitem{Hel12a}
V. Hellemans, P.-H. Heenen and M. Bender,
Phys. Rev. C \textbf{85}, 014326 (2012).

\bibitem{gall94a}
B. Gall, P. Bonche, J. Dobaczewski, H. Flocard, and P.-H. Heenen,
Z. Phys. A \textbf{348}, 183 (1994).

\bibitem{BlaRip}
J. P. Blaizot and G. Ripka,
\emph{Quantum Theory of Finite Systems}
(MIT, Cambridge, MA, 1986).

\bibitem{Ena99a}
K. Enami, K. Tanabe, and N. Yoshinaga,
Phys. Rev. C \textbf{59}, 135 (1999).

\bibitem{baye86a}
D. Baye and P.-H. Heenen,
J. Phys. A\textbf{19}, 2041 (1986).

\bibitem{satula10a}
W. Satu{\l}a, J. Dobaczewski, W. Nazarewicz and M. Rafalski,
Phys. Rev. C \textbf{81}, 054310 (2010).

\bibitem{satula12a}
W. Satu{\l}a, J. Dobaczewski, W. Nazarewicz and T. R. Werner
Phys. Rev. C \textbf{86}, 054316 (2012).

\bibitem{Yu03a}
A. Bulgac and Y. Yu,
Phys. Rev. Lett. \textbf{88}, 042504 (2002).

\bibitem{Bon90a}
P. Bonche, J. Dobaczewski, H. Flocard, P.-H. Heenen, and J. Meyer,
Nucl. Phys. \textbf{A510}, 466 (1990).

\bibitem{Hee93a}
P.-H. Heenen, P. Bonche, J. Dobaczewski, and H. Flocard,
Nucl. Phys. \textbf{A561}, 367 (1993).

\bibitem{avez12a}
B. Avez and M. Bender,
Phys. Rev. C \textbf{85}, 034325 (2012).

\bibitem{sadoudi13b}
J. Sadoudi, T. Duguet, J. Meyer, and M. Bender,
Phys. Rev. C \textbf{88}, 064326 (2013).

\bibitem{raimondi14a}
F. Raimondi, K. Bennaceur, and J. Dobaczewski,
J. Phys. G \textbf{41}, 055112 (2014).



\end{thebibliography}
\end{document}